\documentclass[twocolumn,showpacs,amsmath,amssymb,prb,superscriptaddress,a4paper]{revtex4}

  \usepackage[T1]{fontenc}
  \usepackage[utf8]{inputenc}
  \usepackage[frenchb, english]{babel}
  \usepackage{amsmath}
  \usepackage{xcolor}
  \usepackage{graphicx}
  \usepackage{graphics}
  \usepackage{dcolumn}
  \usepackage{pgf,pgfarrows,pgfautomata,pgfheaps,pgfnodes,pgfshade}
  \usepackage{multimedia}
  \usepackage{braket}
  \usepackage{lmodern}
  \usepackage{picture}
  \usepackage{psfrag,fancyhdr,layout,appendix,subfigure}
  \usepackage{epstopdf}
  \usepackage{placeins}
  \usepackage{bm}
  \usepackage{anysize}
  \usepackage{times,color}
  \usepackage{subfigure}
  \usepackage{mathptmx}
  \usepackage{amssymb}
  \usepackage{amsfonts}
  \usepackage{amsmath}
  \usepackage{pgfplots}

\DeclareGraphicsExtensions{.jpg,.mps,.pdf,.png,.gif}

\begin{document}

\bibliographystyle{plain}

\preprint{APS/123-QED}

\author{L\'{e}onard Monniello}

\affiliation{UPMC Univ Paris 06, UMR 7588, Institut des NanoSciences de Paris, 4 Place Jussieu, F-75005 Paris, France}
\affiliation {CNRS, UMR 7588, Institut des NanoSciences de Paris, 4 Place Jussieu, F-75005 Paris, France}

\author{Catherine Tonin}

\affiliation{UPMC Univ Paris 06, UMR 7588, Institut des NanoSciences de Paris, 4 Place Jussieu, F-75005 Paris, France}
\affiliation {CNRS, UMR 7588, Institut des NanoSciences de Paris, 4 Place Jussieu, F-75005 Paris, France}

\author{Richard Hostein}

\affiliation{UPMC Univ Paris 06, UMR 7588, Institut des NanoSciences de Paris, 4 Place Jussieu, F-75005 Paris, France}
\affiliation {CNRS, UMR 7588, Institut des NanoSciences de Paris, 4 Place Jussieu, F-75005 Paris, France}

\author{\\Aristide Lemaitre}

\affiliation{CNRS, UPR 20, Laboratoire de Photonique et Nanostructures, 
Route de Nozay, F-91460 Marcoussis, France}

\author{Anthony Martinez}

\affiliation{CNRS, UPR 20, Laboratoire de Photonique et Nanostructures, 
Route de Nozay, F-91460 Marcoussis, France}

\author{Valia Voliotis}
\email{voliotis@insp.jussieu.fr}

\affiliation{UPMC Univ Paris 06, UMR 7588, Institut des NanoSciences de Paris, 4 Place Jussieu, F-75005 Paris, France}
\affiliation {CNRS, UMR 7588, Institut des NanoSciences de Paris, 4 Place Jussieu, F-75005 Paris, France}

\author{Roger Grousson}

\affiliation{UPMC Univ Paris 06, UMR 7588, Institut des NanoSciences de Paris, 4 Place Jussieu, F-75005 Paris, France}
\affiliation {CNRS, UMR 7588, Institut des NanoSciences de Paris, 4 Place Jussieu, F-75005 Paris, France}

\title{Excitation-induced dephasing in a resonantly driven InAs/GaAs quantum dot}

\begin{abstract}

We report on coherent emission of the neutral exciton state in a
single semiconductor self-assembled InAs/GaAs quantum dot embedded
in a one-dimensional waveguide, under resonant picosecond pulsed excitation.
Direct measurements of the radiative lifetime and coherence time are
performed as a function of excitation power and temperature. The
characteristic damping of Rabi oscillations which is observed, is
attributed to an excitation-induced dephasing due to
a resonant coupling between the emitter and the acoustic phonon bath
of the matrix. Other sources responsible for the decrease of the coherence time have been
evidenced, in particular an enhancement of the radiative
recombination rate due to the resonant strong coupling between the dot and the one-dimensional optical mode. As a
consequence, the emission couples very efficiently into the waveguide mode leading to an additional relaxation term
of the excited state population.

\end{abstract}

\maketitle

Important research has been devoted the past decade in understanding
the mechanisms responsible for dephasing in semiconductor quantum
dots (QD). Indeed, coherence is a key issue to address if using QDs as qubits to ensure for instance, high fidelity
operations in coherent control schemes \cite{ref1} or obtain a single-photon source with a high degree of indistinguishability
\cite{ref2}. Because of their strong interaction with their environment, 
the coherence time is not radiatively limited as expected theoretically 
and pure dephasing processes can occur depending on the kind of 
quantum dots and the nature of their host matrix \cite{ref3}.
This can be well understood in the case of incoherent optical pumping of the dot, when using for instance non-resonant excitation
leading to coupling to continuum states in the wetting layer \cite {QQWang, Vasanelli, VillasBoas}, or to multiexcitonic transitions
\cite{multiexciton}.
Strictly resonant pumping of a given optical transition in the dot appears then as the most reliable way to keep the coherence of the state. 
However, only a few demonstrations have been reported in the literature due to the technical difficulties in performing resonant 
excitation experiments. Recently, resonant excitation configurations have been achieved allowing the observation of the resonant
luminescence of a single QD \cite{Muller, Melet, HaiSon, Ates, Matthiesen}. Other techniques, like differential transmission \cite{Stievater},
four-wave mixing\cite{Borri} or photocurrent detection \cite{Stufler, Ramsay}, allowed also the resonant manupilation of a single QD.
Even though, important dephasing occurs observed both in cw and pulsed excitation regimes,
showing a spectral broadening of the Mollow triplet \cite{Ulrich} or a damping of Rabi oscillations \cite{Borri, Melet, Ramsay} respectively. The
damping has been shown to be power dependent \cite{Stievater, QQWang, Melet, Ramsay, Mogilevtsev} and the main mechanism for such an excitation
induced dephasing (EID) is the interaction with longitudinal acoustic (LA) phonons that constitute an intrinsic limitation for coherence. Numerous
theoretical approaches have been used to investigate the effect of phonon interactions on the coherent manipulation of excitons in QDs, using exact
or perturbative methods \cite{Forstner, Machnikowski, Nazir, Krummheuer}, path integral formalism \cite{Vagov} or polaron transform
\cite {McCutcheon, Roy}. For temperatures below 30 K which are the usual experimental conditions where single dot spectroscopy is carried out,
similar predictions have been found \cite {McCutcheon} and the weak-coupling regime is enough to describe the physical situation in a first approximation. 

In this Letter, we report on resonant luminescence from single QD neutral exciton under picosecond (ps) pulsed excitation. Direct measurements of
the radiative lifetime $T_{1}$ and coherence time $T_{2}$ are reported as a function of temperature and excitation power in order to investigate the
different EID mechanisms taking place. The role of phonons in dephasing is definitely dominant and we choose the weak coupling picture for the dot-LA
phonons interaction in order to explain the observed damping of the Rabi oscillations \cite{Ramsay, McCutcheon}. A quadratic frequency dependent behavior
is found which is characteristic of a resonant coupling between the two-level system and the phonon bath \cite {Machnikowski, Nazir}.
Moreover, depending on the specific QDs growth conditions and geometry (QDs embedded in micropillars, nanowires, photonic crystal waveguides, etc...)
a shaping of the electromagnetic spectral density can give rise to a modification of the spontaneous emission rate
\cite{micropillars, Press, nanowires, PhC} and influence the coherence properties of the system. In our samples structure, the dots are coupled to a
single-mode 1D waveguide (WG) that modifies the spontaneous emisssion rate \cite{Domokos} and leads to an additional relaxation of the population.
The coupling to the optical mode is different from one dot to another because of the random distribution of the dots in the InAs layer and thus, is
dot position dependent. Two typical results will be presented: for one kind of dots (hereafter called of \textit{type A}) that are not efficiently
coupled to the WG mode and for another one (QDs of \textit{type B}) where on the contrary the coupling is very efficient.

InAs/GaAs self-assembled QDs, were grown by MBE on a planar $[001]$
GaAs substrate. In order to address the fundamental exciton state on
resonance, the dots are embedded in a two-dimensional GaAlAs
transverse single-mode waveguide \cite{MeletSM}. Micrometer
ridges are etched on the top surface in order to reduce the volume
of the optical mode and enhance the light-matter interaction. In contrast to the case of dots embedded
in microcavities there is no need here to match the cavity mode energy with the QD emission in order to achieve the strong coupling regime. The QDs
are excited by ps pulses provided by a tunable mode-locked Ti:sapphire laser and focused on the sample using a microscope objective. The sample is
fixed on the cold finger of a three-axes optical cryostat (7 K) specifically designed for the waveguiding geometry. More experimental details can
be found in Refs [9] and [32].
In this geometry the laser propagates in the dots plane and the
single QD luminescence is collected from the ridge top surface by a
confocal  micro-photoluminescence ($\mu PL$) detection setup. Since
the laser beam is confined in the guided mode, the scattered light
is greatly suppressed and at low pump power the resonant
luminescence is almost laser background-free. Using this geometrical
configuration, resonant Rabi oscillations of the exciton state in a
single QD have been observed \cite{Melet}. Coherent control
experiments have also been achieved and allowed determining the main
decoherence mechanisms in the system \cite{Enderlin}.

\begin{figure}[t!]
\begin{center}
\includegraphics[width=7cm,height=5cm]{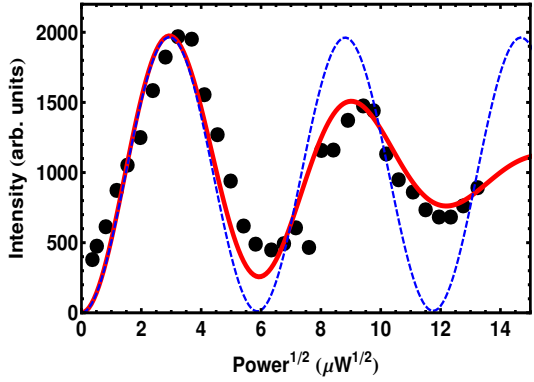}
\end{center}
\caption{Resonant Rabi oscillation of the population in one
particular QD of \textit{type A} at 7 K: the black dots correspond to the emission
intensity of the exciton plotted as a function of the square root of
the pump power. The blue dotted curve is a fit using the bare optical
Bloch equations and the red one includes coherence relaxation (see text).}
\label{oscrab}
\end{figure}

The non-linear interaction between the two-level system (TLS) and
the resonant field gives rise to the well-known Rabi oscillation
(RO) of the excited level population. In pulsed excitation, the
emission intensity oscillates as a function of the pulse area which
is proportional to the square root of the incident laser power \cite{MW}. A typical RO
of a QD emitting at 930.3 nm is shown in Fig. 1. Two main features
appear on this curve, the rapid damping and the limited number of oscillations that can be recorded. The former behaviour is related, as it will be
discussed in the following, to the excitation-induced non-linear coupling between the optically driven dipole and the phonon modes.
The latter problem is due in fact to the residual scattered laser
which becomes important when the pump power is increased. This is one
of the main difficulties when studying the resonant PL as compared
to photocurrent measurements  where numerous oscillations can be observed \cite{Stufler, Ramsay}. Due to the finite
radiative lifetime and the limited coherence time of the system a
damping of RO is expected. We have measured by time-resolved PL the on-resonance lifetime $T_{1}$ for this dot and found $800 ps$. The coherence time
$T_{2}$ has also been measured by coherent control experiments and found to be of the order of $T_{1}/2$. Nevertheless, using the standard optical
Bloch equations for a TLS with constant $T_{1}$ and $T_{2}$ is not enough to accurately simulate the experimentally observed damping, as shown in
Fig. 1 by the blue dotted curve. A power dependent dephasing has to be introduced in the model in order to explain the rapid damping of RO. For
resonantly driven QDs, we expect that the coupling to phonons should be the most relevant dephasing process
\cite{Forstner, Machnikowski, Nazir, Krummheuer, Vagov, McCutcheon, Roy}. The RO behaviour can then
be more accurately adjusted as shown in Fig.1 by the red curve. 

\begin{figure}[t!]
\begin{center}
\includegraphics[width=7cm,height=5cm]{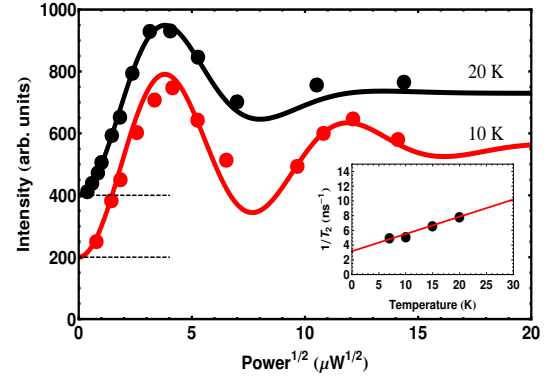}
\end{center}
\caption{Rabi oscillation of a \textit{type A} QD at 10 K and 20 K in red and black curves respectively. The inset shows the linear dependence of
the coherence time $T_{2}$ with temperature for excitation with $\pi/2$ pulses.}
\label{rabtemp}
\end{figure}

\begin{figure}[t!]
\begin{center}
\includegraphics[width=7cm,height=5cm]{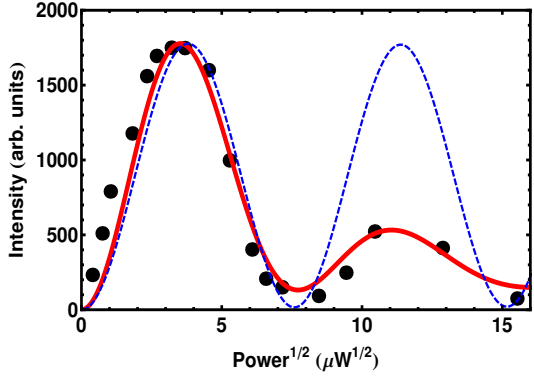}
\end{center}
\caption{Resonant RO of the population in a \textit{type B} QD: the black dots correspond to the emission
intensity of the exciton plotted as a function of the square root of
the pump power. The blue dotted curve is a fit using the bare optical
Bloch equations while the red one includes excitation induced coherence and population relaxation.}
\label{oscrabibi}
\end{figure}

Thereafter, we performed $\mu PL$
experiments with resonant ps pulses by varying the sample
temperature from 7 to 30 K. Above 30 K, the resonant emission
intensity becomes too weak to be observed. 
In Fig.2 we show two RO curves at 10 and 20 K for another dot of \textit{type A}. As expected, the damping has increased with temperature. To
identify whether the interaction with acoustic phonons is the only
EID mechanism, we modeled the exciton - LA phonon coupling using a
perturbation approach {\cite{Ramsay, Forstner, Machnikowski, Mogilevtsev}. The decoherence rate is
found to be proportional to the square of the bare Rabi frequency $\Omega_R$,
which is characteristic of an EID process. The optical Bloch equations take then
the form:

\begin{equation}
\left\{
\begin{array}{ll}
\dot{\sigma}_{11}(t) = - \dfrac{i}{2} \ \Omega_R (t) \ \left(\hat{\sigma}_{10}(t)-\hat{\sigma}_{01}(t)\right)
 - \dfrac{\sigma_{11}(t)}{T_1} \vspace{2mm} \\
\dot{\hat{\sigma}}_{01}(t) =  \dfrac{i}{2} \ \Omega_R (t) \
\left(\sigma_{11}(t)-\sigma_{00}(t)\right)\\-
\left(\dfrac{1}{T_2}+\kappa \left(T, \Omega_{R}(t)\right) \right)
\hat{\sigma}_{01}(t)
\end{array}
\right.
\end{equation}

$T_{1}$ is determined experimentally and is kept constant. $\kappa \left(T, \Omega_{R}(t)\right)$ is the damping rate due to phonon coupling which
takes the simple form for temperatures above 10 K :

\begin{equation}
\kappa(T,\Omega_{R}(t)) \approx \textit{K}\ T \Omega_R^2(t) = \dfrac{(\mathcal{D}_e-\mathcal{D}_h)^2}{4\pi\hbar^2 \mu
c_s^5} \ k_BT \Omega_R^2(t)
\end{equation}

The proportionality coefficient \textit{K} depends only on the properties of the surrounding GaAs bulk material that can be found in the literature
{\cite{Krummheuer}: the effective band-gap deformation potential $(\mathcal{D}_e-\mathcal{D}_h)\approx-8.5$ eV, $\mu\approx 5.4$  $g.cm^{-3}$, the
sound velocity $c_S\approx5110$  $m.s^{-1}$. We can then estimate $\textit{K}_{theo} \approx10$ $fsK^{-1}$. In fact, at temperatures lower than 10 K the
linear dependence is not valid anymore because $k_BT\leq \hbar\Omega_R/2$ \cite{Ramsay, McCutcheon}. By adjusting numerically the RO curves at different
temperatures with eqs (1), we find an experimental value of $\textit{K}_{exp} \approx 25 \ fsK^{-1}$ which agrees with the above estimation. The coherent
control measurements as a function of temperature allowed to determine the coherence time $T_{2}$ temperature dependence and a linear variation has been
found in agreement with the model (see inset in Fig. 2). When extrapolating at zero temperature, it appears that the coherence time is not radiatively
limited. Indeed, $T_{1}$ has been measured to be $600$ ps for this dot but at zero temperature $T_{2}\approx 330 ps\neq 2T_{1}$. Thus, it
seems that another source of pure dephasing exists with a rate comparable to the coupling to phonons of the order of 2.2 $ns^{-1}$. This 
additional dephasing has been discussed by different groups \cite{Besombes,Berthelot,Nguyen} and is likely due to the dot fluctuating electrostatic
environment which is a consequence of charge trapping in the vicinity of the dot. This supplementary mechanism has been evidenced only for $20\%$ of
the studied dots showing again the influence of the dot specific environment.
In the case of EID due to phonons, the luminescence intensity $\mathcal{L}$ tends to the limiting value $\mathcal{L}_{\infty} = \mathcal{L}_{max}/2$
for high pump power corresponding to the stationary occupation of one half \cite{MW}. This is the case for the dots of type A like 
shown in Figs 1 and 2. In our experiments neither a renormalization of the Rabi frequency with increasing temperature \cite{Ramsay} was observed,
nor a revival of the RO with higher pulse area, as predicted theoretically by Vagov and coworkers \cite{Vagov}. The reasons for that may be due to
experimental limitations because the luminescence becomes too weak to be detected above $ 30 K$ and the resonant laser completely blurs the emission
of the dots for large pulseareas.

Another typical trend that we found is that at high pump power $\mathcal{L}_{\infty} <\mathcal{L}_{max}/2$, and even $\mathcal{L}_{\infty}\rightarrow 0$
in some cases (Fig. 3). These dots are the one denoted as \textit{type B}. In this situation, the excitation induced pure dephasing is not
the only mechanism explaining the damping and an additional power dependent relaxation term for the population has to be taken into account.
This would be equivalent to a term that enhances the emission rate as a function of the pump power. A similar behaviour has been reported
in previous resonant experiments in thickness fluctuations interface dots \cite{Melet} embedded also in 1D waveguides. We have corroborated
this result by performing resonant time-resolved $\mu PL$ experiments as a function of the pump power. Indeed, for \textit{type B}
QDs the radiative lifetime gets shorter as the pump power
increases, whereas in the case of \textit{type A} QDs the radiative lifetime does not
vary significantly with power. Fig. 4 shows the time-resolved $\mu PL$ of the exciton in a \textit{type B} QD
excited on resonance with $\pi$, $3\pi/2$, and $5\pi/2$ pulses. The
radiative lifetime is reduced from 800 ps to 540 ps respectively.
The inset in Fig. 4 shows that the lifetime is inversely proportional to the
pump power thus, to the square of the Rabi frequency. This additional relaxation term has been added phenomenologically in the Bloch equations
and fits quite well the oscillation curve (Fig. 3). They read now:

\begin{equation}
\left\{
\begin{array}{ll}
\dot{\sigma}_{11}(t) =- \dfrac{i}{2}\Omega_R (t)
\left(\hat{\sigma}_{10}(t)-\hat{\sigma}_{01}(t)\right)\\*
  - \left(\dfrac{1}{T_1}+\alpha \ \Omega_R^2(t) \right) \ \sigma_{11}(t) \vspace{2mm} \\
\dot{\hat{\sigma}}_{01}(t) = \dfrac{i}{2}\Omega_R (t)
\left(\sigma_{11}(t)-\sigma_{00}(t)\right)\\
 - \left(\dfrac{1}{T_2}+ K\ T \ \Omega_R^2(t) \right) \ \hat{\sigma}_{01}(t)
\end{array}
\right.
\end{equation}

$\alpha$ is an adjustable parameter used to fit the power dependence of the lifetime $T_{1}$. We believe that it is specific to the coupling between
the dipole emission and the 1D WG mode. The WG modifies the structure of the electromagnetic environment, changes the emission properties of the dipole
which is inside and as a consequence modifies the emission rate \cite{Domokos,Lecamp}. This modification is not uniform for all the optical modes but
occurs only with the WG resonant mode. We can assume in a first approximation that the emission rate to all other radiative modes remains unchanged.
If we define the fraction of spontaneously emitted light into a given mode, the so-called coupling factor $\beta$, we can write the total
\textit{resonant} emission rate $\gamma_{R}=1/T_{1}$ as:
\[
	\gamma_{R}=(1-\beta)\gamma_{0}+\beta(\gamma_{0}+\gamma_{wg})=\gamma_{0}+\beta\gamma_{wg} (4)
\]
$\gamma_{0}$ is the spontaneous emission rate in the sample and $\gamma_{wg}$ is the modified emission rate due to the resonant coupling to the 1D
optical mode.  

\begin{figure}[t!]
\begin{center}
\includegraphics[width=7cm,height=5cm]{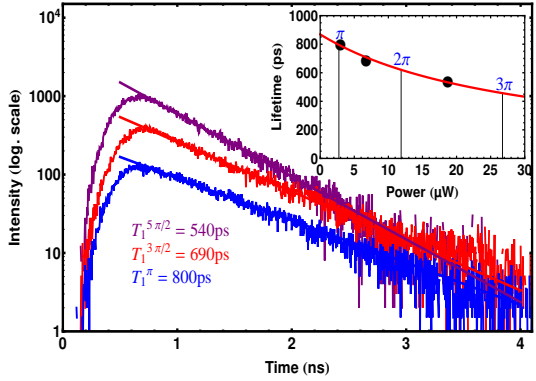}
\end{center}
\caption{Time-resolved $\mu PL$ of the \textit{type B} QD excited on resonance for different pulse areas. The inset shows that the radiative lifetime
scales like the inverse of the pump power. }
\label{tpspuis}
\end{figure}

As a consequence the modified emission rate $\gamma_{wg}$ will be power dependent. To show it qualitatively we may use the results of cavity-QED in
the strong coupling regime \cite{Auffeves, Cui, Press}. Although our QDs are not coupled to cavity photons, they are resonantly coupled with those
of the 1D single-mode WG in the non-linear Rabi regime. The modification of the emission rate, allows us to define an enhancement factor $F$ which
is the ratio between $\gamma_{wg}$ and $\gamma_{0}$, $F$ being analogous to the well-known Purcell factor $F_{p}$ \cite {Purcell}. In the strong
coupling regime, the generalized Purcell factor reads: 
$F_{p}=\frac{4g^{2}}{\gamma_{c}\gamma_{0}}$,
with $\gamma_{c}$ the intracavity decay rate and $g$ the light-matter coupling \cite{Cui,Press}. The strong coupling regime holds when
$g>\gamma_{c},\gamma_{0}$. In pulsed experiments, the relevant parameter equivalent to $1/\gamma_{c}$ is the pulse width $\tau$ equal to $2 ps$ in
our case. The coupling parameter $g$ is related to the Rabi frequency 
$\Omega_{R}$ thus, to the number of photons $N$ $(N>>1)$ in the pulse \cite{Domokos}: $(2g)^{2}= \Omega_{R}^{2}= N(2g_{0})^{2}$, $g_{0}$ being the
single photon coupling. 
Therefore $\gamma_{wg}=N(2g_{0})^{2}\tau$, and the modified rate scales like the pump power, which is exactly what we observe experimentally.
The coupling factor $\beta$ can be estimated using the factor $\alpha$ value (eq. 3) and eq. (4). $\beta$ is characteristic of the WG structure
and in the case of $1\mu m^{2}$ ridges dimensions, we find a few percent ($3\%$ in the case of the studied dot). Achieving a large $\beta$ factor
by reducing the WG cross-section dimensions would be of great interest for the development of very efficient single photon sources. Although in
our etched structures it is rather difficult to reduce the ridge dimensions, this can be obtained in other geometries like in nanowires \cite{nanowires}
or photonic crystal waveguides \cite{PhC} where very large $\beta$ coupling factors up to $95 \%$ have been reached. 
In our sample structure the fact that $\gamma_{wg}$ varies from dot to dot is related to the light-matter coupling $g_{0}$, which depends mainly on
the position of the dot with respect to the maximum of the field amplitude. Therefore, reducing the dimensions of the WG would have as a consequence
to enhance the resonant light-matter coupling due to the larger overlap between the dot absorption cross-section and the cross-section of the optical
mode.

We have presented two extreme cases of dots, \textit{type A} and \textit{B} with
$\mathcal{L}_{\infty} \rightarrow \mathcal{L}_{max}/2$ and
$\mathcal{L}_{\infty} \rightarrow 0$ respectively. However, different situations
have been encountered with $\mathcal{L}_{\infty}$ ranging between
these two values. For all dots, the damping rate $K$ caused by
phonons is of the same order of magnitude whereas the population
relaxation rate $\alpha$  is related to the modification of the emission rate and thus dot-dependent. The resonant coupling to the 1D WG mode is
therefore responsible for the differences observed in the oscillations damping.

In summary, we have shown the dominant role of acoustic phonons in
the EID processes that inherently limit the coherence of the system. The
damping rate $\textit{K}$ is similar for all the
studied dots and of the same order of magnitude as the theoretically
expected for GaAs. The linear temperature dependence of the
measured coherence time supports this result. For a certain number of dots well coupled to the 1D waveguiding mode an additional damping is observed
related to an excitation-induced relaxation of the population. The spontaneous emission rate enhancement scaling with the pump power acts as a leak
of population that can lead in certain cases to a complete vanishing of the resonant luminescence at high excitation. Further investigation with
longer than $2 ps$ pulses is currently carried out, since interesting effects have been predicted theoretically. In particular, a low quality of
Rabi oscillations has been calculated for ultrashort pulses while a decrease of the damping has been predicted with longer pulses
\cite{Forstner, Machnikowski}. This could be a way to minimize the EID and achieve more reliable coherent operations.

The authors acknowledge financial support from the French Agence Nationale de
la Recherche (ANR-11-BS10-010) and the C'Nano Ile-de-France (\no 11017728).







\end{document}